# Tunable superconducting nanoinductors

**Anthony J Annunziata[1], Daniel F Santavicca[1], Luigi Frunzio[1], Gianluigi Catelani[2], Michael J Rooks[1], Aviad Frydman[3], and Daniel E Prober[1,2]**

[1]Department of Applied Physics, Yale University, New Haven, CT 06511
[2]Department of Physics, Yale University, New Haven, CT 06511
[3]Department of Physics, Bar-Ilan University, Ramat Gan 52900, Israel

E-mail: anthony.annunziata@yale.edu, daniel.prober@yale.edu

**Abstract.** We characterize inductors fabricated from ultra-thin, approximately 100 nm wide strips of niobium (Nb) and niobium nitride (NbN). These nanowires have a large kinetic inductance in the superconducting state. The kinetic inductance scales linearly with the nanowire length, with a typical value of 1 nH/μm for NbN and 44 pH/μm for Nb at a temperature of 2.5 K. We measure the temperature and current dependence of the kinetic inductance and compare our results to theoretical predictions. We also simulate the self-resonant frequencies of these nanowires in a compact meander geometry. These nanowire inductive elements have applications in a variety of microwave frequency superconducting circuits.

## 1. Introduction

Inductors are ubiquitous in high frequency circuits. Most inductors utilize the magnetic self-inductance of a straight wire or coil. Magnetic self-inductance is a consequence of Faraday's law, and depends on the energy stored in the magnetic field due to a current. The magnetic self-inductance of a straight wire scales with the length of the wire, but depends only logarithmically on the wire diameter, and is typically ~1 pH/μm.[1] Hence, it is difficult to make a large magnetic inductance in a compact planar geometry, as is desired for many emerging high frequency micro- and nano-circuit applications.

The kinetic inductance ($L_K$) does not arise from the energy stored in the magnetic field but rather from the kinetic energy stored in the motion of the charge carriers. In a nanowire, $L_K$ scales with the wire length and inversely with the cross-sectional area in both normal metals and superconductors. Because of this, $L_K$ can be very large in a wire of moderate length (~ 100 μm) with a nanoscale cross-section. In a non-superconducting nanowire, the resistive component of the impedance dominates the kinetic inductive component of the impedance at frequencies up to ~THz. In a superconductor, however, the dc resistance is zero and the impedance from dc to GHz frequencies can be dominated by the kinetic inductance of the supercurrent.

An alternative approach to achieving a large inductance at GHz frequencies is based on an array of Josephson junctions.[2-4] Compared to nanowires, this requires a relatively complex fabrication process.

Furthermore, the non-linearity of the Josephson inductance with current, while useful for some devices, is not desirable in all applications. Superconducting nanowires offer a large inductance that depends only weakly on current in a geometry that is relatively easy to fabricate. This makes them attractive for use in a variety of superconducting microwave circuits, including photon detectors,[5-7] metamaterials,[8] low-loss compact filters,[9] and quantum bits.[4,10]

Previous studies of kinetic inductance in superconductors have mainly focused on larger, micron-scale structures.[11-14] While easier to fabricate, these structures have considerably smaller inductance per unit length than a wire with nanoscale cross-section. The additional capacitance of these two-dimensional structures also results in a lower self-resonant frequency. Reference 10 discusses the concept of using a niobium nitride (NbN) nanowire as an inductive shunt to minimize low-frequency charge noise in a superconducting quantum bit. Previous work by the same author reported the current dependence of kinetic inductance in NbN nanowires at low temperature.[15] We expand on the studies of reference 15 by characterizing both the current and temperature dependence of NbN nanowires, and also by studying niobium (Nb) nanowires. While NbN nanowires have greater inductance per unit length, Nb nanowires may be preferable for some applications due to ease of fabrication. We compare our experimental results to predictions from both Ginzburg-Landau and BCS theories.

The kinetic inductance of a superconducting nanowire can be calculated by equating the total kinetic energy of the Cooper pairs with an equivalent inductive energy: $n_s(lwd)mv^2 = ½L_KI^2$, where $I$ is the supercurrent in the nanowire, $n_s$ is the density of Cooper pairs, $lwd$ is the volume of the conductor (length × width × thickness), $2m$ is the mass of a Cooper pair, and $v$ is the average velocity of the Cooper pairs. This yields $L_K = (m/2n_se^2)(l/wd)$.

Within the Ginzburg-Landau (GL) theory, the temperature dependence of the Cooper pair density can be expressed as $n_s(T) \approx n_s(0)(1\text{-}T/T_c)$.[16] Solving for $L_K(T)$ for small currents, we obtain

$$L_K(T) = \frac{m}{2e^2}\left(\frac{l}{wd}\right)\left(\frac{1}{n_s(T)}\right) \approx \frac{m}{2e^2}\left(\frac{l}{wd}\right)\left(\frac{1}{n_s(0)\left(1-T/T_c\right)}\right) = L_K(0)\left(\frac{1}{1-T/T_c}\right). \quad (1)$$

This can be expressed in terms of the GL magnetic penetration depth $\lambda$, where $\lambda^2(T) = m/(2\mu_0 n_s(T)e^2)$,[16] with $\mu_0$ the permeability constant:

$$L_K(T) = \mu_o\lambda^2(T)\left(\frac{l}{wd}\right). \quad (2)$$

The approximation of equation (1) applies in one dimension for $T$ near $T_c$ and $I \approx 0$.[16]

In addition to depending on temperature, $n_s$ also depends on current. In general, this dependence is difficult to obtain analytically. Using GL theory, and following the results in reference 16, we express $n_s$ and $I$ in terms of $k\xi$, where $k$ is the gradient of the phase along the length of the superconducting nanowire and $\xi$ is the superconducting coherence length. We then calculate the current dependence of $L_K$ by numerically solving this system of equations for $L_K(I/I_c)$:[17]

$$L_K(k\xi) = \frac{m}{2e^2}\left(\frac{l}{wd}\right)\left(\frac{1}{n_s(k\xi)}\right) \approx \frac{m}{2e^2}\left(\frac{l}{wd}\right)\left(\frac{1}{n_s(0)\left(1-3k^2\xi^2\right)}\right) = L_K(0)\left(\frac{1}{\left(1-3k^2\xi^2\right)}\right) \quad (3)$$

$$\frac{I}{I_c} \approx \frac{3\sqrt{3}}{2}\left(k\xi - k^3\xi^3\right) \quad (4)$$

where $I_c$ is the GL critical current for pair-breaking.[16] Solving these equations by series expansion in the limits of small and large current yields results that agree with those found in reference 14. We note that these GL results are only strictly valid for $T \approx T_c$.

We can also solve for $L_K$ using BCS theory, which is valid for all $T$. In the low-frequency limit ($hf << k_BT$), the Mattis-Bardeen formula for the complex conductivity can be written in terms of the ratio of the imaginary conductivity $\sigma_2$ to the normal state conductivity $\sigma_n$ as[16]

$$\frac{\sigma_2}{\sigma_n} = \frac{\pi\Delta}{hf}\tanh\left(\frac{\Delta}{2k_B T}\right) \tag{5}$$

where $\Delta = \Delta(T)$ is the temperature-dependent superconducting energy gap (we assume here zero bias current). The imaginary component of the impedance is due to kinetic inductance, and hence we can write equation (5) as

$$L_K = \left(\frac{l}{w}\right)\frac{R_{sq}h}{2\pi^2\Delta}\frac{1}{\tanh\left(\frac{\Delta}{2k_B T}\right)} \tag{6}$$

where $R_{sq}$ is the sheet resistance in the non-superconducting state. It is straightforward to solve this for $T \ll T_c$ using $\Delta(0) \approx 1.76kT_c$, and for $T \approx T_c$ using $\Delta(T) \approx \Delta(0)1.74(1\text{-}T/T_c)^{1/2}$.[16] Outside of these limiting cases, $\Delta(T)$ must be solved numerically, as described in reference 16.

The dependence of $\Delta$ on the bias current is more complex, and for simplicity we consider here only the zero temperature limit, for which, from equation (6), we can write

$$\frac{L_K(I)}{L_K(0)} = \frac{\Delta_{00}}{\Delta} \tag{7}$$

where $\Delta_{00}$ is the energy gap at zero temperature and zero bias current. In the dirty limit (the mean free path much shorter than the coherence length), we can express $\Delta$ and the bias current density $j$ as functions of the phase gradient $k$:[18]

$$\Delta = \Delta_{00}e^{-\pi\zeta/4} \tag{8}$$

$$j = 2eN(0)Dk\Delta\left(\frac{\pi}{2} - \frac{2}{3}\zeta\right) \tag{9}$$

where $\zeta = Dk^2/(2\Delta)$, $D$ is the diffusion coefficient, and $N(0)$ is the density of states at the Fermi surface. The critical current density $j_c$ is the maximum of $j(\zeta)$, which occurs for $\zeta = 0.300$:

$$j_c = 1.491eN(0)\sqrt{D}\Delta_{00}{}^{3/2}. \tag{10}$$

We can express the ratio of the bias current $I$ to the critical current $I_c$ as

$$\frac{I}{I_c} = 1.897e^{-3\pi\zeta/8}\sqrt{\zeta}\left(\frac{\pi}{2} - \frac{2}{3}\zeta\right). \tag{11}$$

Thus we can find $\zeta$ for a particular $I/I_c$, which can be used in equation (8) to find $\Delta$ and hence $L_K(I)$. Just below $I_c$, $L_K(I)$ reaches a maximum value of $1.266L_K(I=0)$.

## 2. Methods and Procedure

The fabrication procedures for the Nb and NbN nanowires studied in this work have been reported in reference 7 and reference 19, respectively. Briefly, Nb or NbN was sputter deposited on an R-plane sapphire substrate. Electron-beam lithography was used to define a single-layer of PMMA that had been spin coated on the surface. The patterned PMMA was used as an etch mask for reactive ion etching of the thin Nb or NbN film. Using this method, Nb nanowires were patterned into planar strips with thickness $d \approx 15$ nm,[20] width $w = 100$ nm, and with a sheet resistance of $R_{sq} \approx 20$ Ω/square at temperatures just above $T_c$. NbN nanowires were patterned into strips with $d \approx 5$ nm, $w = 130$ nm, and $R_{sq} \approx 900$ Ω/square just above $T_c$. Typically, these nanowires are patterned in a square meander geometry to conserve space and to minimize the magnetic mutual-inductance between strips. In figure 1, scanning electron micrographs of two devices are seen: (a) a 10×10 μm$^2$ Nb meander with total length $l$ = 500 μm and $L_K \approx$ 22 nH at 2.6 K; and (b) a 2×2 μm$^2$ meander with total length $l$ = 22 μm and $L_K \approx$ 1 nH at 2.6 K. In general, NbN nanowires can be made thinner than Nb nanowires while preserving a relatively high critical temperature, in part because of the shorter coherence length in NbN compared to Nb.

The inductance is measured by incorporating the nanowire meander into a lumped-element resonant circuit, as seen in figure 1(c). The device is connected in parallel with a chip capacitor with a capacitance $C_o$ and a chip resistor with $R_L$ = 50 Ω. The value of $C_o$ is either 47 pF or 100 pF, chosen to ensure that the resonant frequency for each device lies within our measurement range of 10-100 MHz. $C_o$ and $R_L$ were each measured at room temperature and at low temperature to ensure their values are temperature-independent. Both the chip resistor and capacitor are located close to the nanowire device (< 5 mm) to ensure that the phase difference introduced by the separation is small. The magnetic inductance of the nanowire, ~ 1 pH/μm, is at least 40 times smaller than $L_K$, and hence is neglected. We estimate the inductance of the wirebonds used to make electrical connection between the device and the sample holder to be ≲ 1 nH, which is also small compared to the kinetic inductance of the devices studied. The kinetic inductance is determined by measuring the resonant frequency $f_o$ and solving for $L_K$ from

$$f_o = \frac{1}{2\pi\sqrt{L_K C_o}} . \tag{12}$$

At the resonant frequency, the inductive and capacitive contributions to the impedance cancel, and the resonator has an impedance of $R_L$ = 50 Ω. Hence, on resonance, the reflection coefficient is a minimum.

A 50 Ω coaxial transmission line couples the resonator to a bias tee. The bias tee separates the dc current and the high frequency probe signal (10-100 MHz). The dc bias current is provided from a low-noise voltage source in series with a large bias resistor, $R_b$ = 1 MΩ. Low-pass filters (LP1 and LP2) are used to filter noise from the dc bias line. The reflection coefficient of the resonant circuit is measured as a function of frequency from 10-100 MHz using a network analyzer (HP3589A), which is coupled to the device via the coupled port of a -20 dB directional coupler. Although the intended applications of these inductors are in the GHz range, the measurements were performed at lower frequencies to mitigate the effects of parasitic reactance in the measurement circuit. The power coupled to the resonant circuit in all measurements was -100 dBm. The reflected power is amplified by low noise amplifiers and measured by the network analyzer. The resonant frequency is identified by a minimum in the reflected power versus frequency. As an example, we plot the resonance curves measured for a 5 x 5 μm$^2$ NbN nanowire meander at four different bath temperatures (2.5, 5.0, 7.0, and 7.5 K) and with zero bias current. The resonance quality factor is limited by the circuit rather than by losses in the nanowire, except for $T$ close to $T_c$. Although the present work does not probe the intrinsic losses in the nanowire, we note that for $T << T_c$ and for frequencies $f << 2\Delta/h$, these devices should be very low loss,[16] which is important for many of the potential applications.

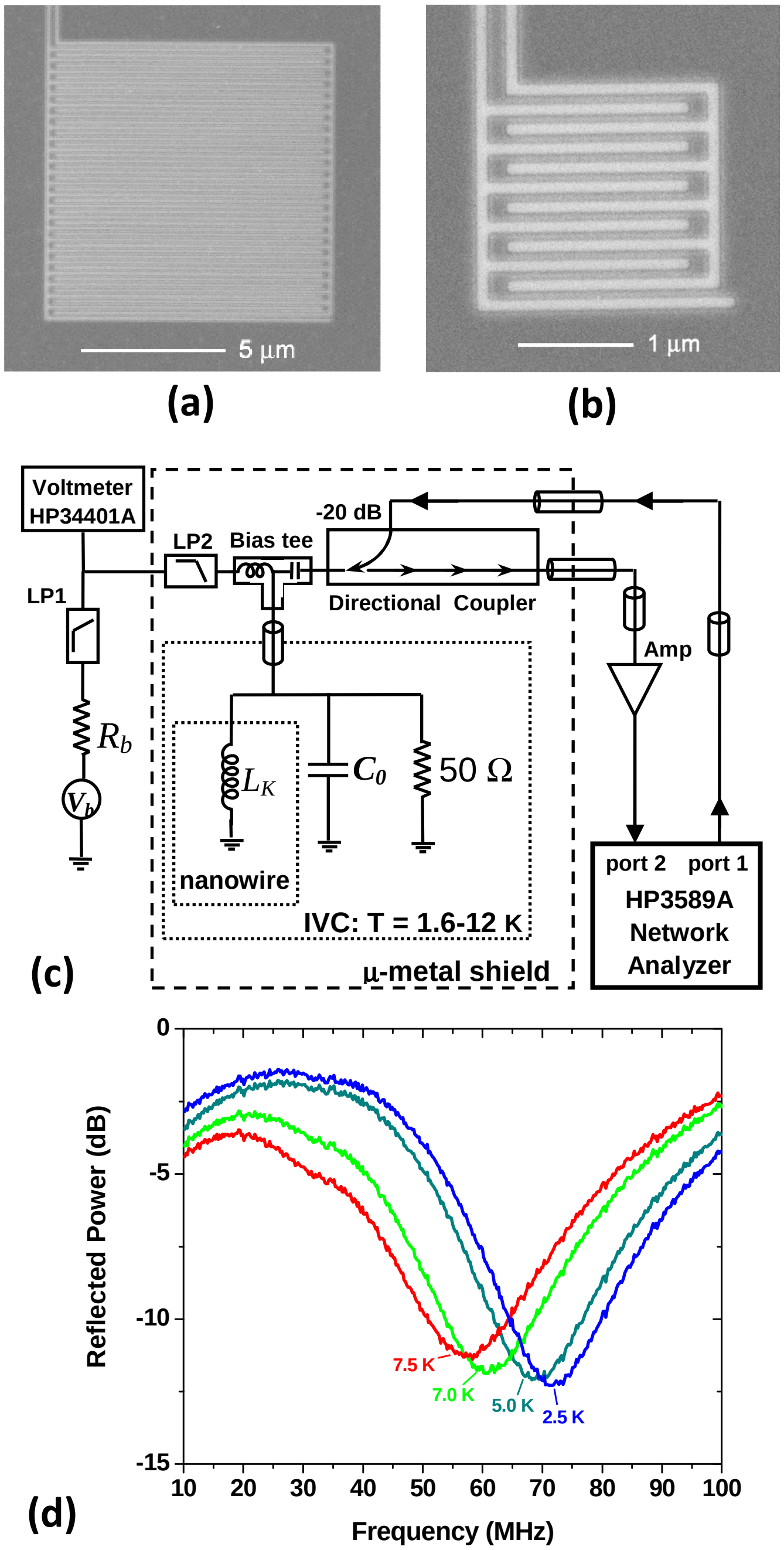

Figure 1. (a) Scanning electron micrograph of a Nb nanowire (darker color is Nb; lighter color is sapphire) in a 10 × 10 $\mu m^2$ meander with total nanowire length $l$ = 500 µm. (b) 2 × 2 $\mu m^2$ Nb meander with total length $l$ = 22 µm. (c) Schematic of measurement setup. (d) Measured resonance curves for 5 x 5 $\mu m^2$ NbN nanowire meander with zero bias current at bath temperatures of 2.5, 5.0, 7.0, and 7.5 K. The value of $C_0$ in this measurement is 47 pF, and the slight decrease in reflected power at low frequency is due to the bandwidth of the directional coupler.

## 3. Results

We report measurements of the temperature and current dependence of the kinetic inductance in a 10 x 10 $\mu m^2$ Nb nanowire meander with $d$ = 15 nm, $w$ = 100 nm, and $l$ = 500 μm (device A) and a 5 x 5 $\mu m^2$ NbN nanowire meander with $d$ = 5 nm, $w$ = 130 nm, and $l$ = 105 μm (device B). We have measured several other devices and verified that the kinetic inductance scales with the total nanowire length, as expected.

In figure 2(a) a plot of $L_K$ versus $T$ is shown for device (A), while in figure 2(b) a plot is shown for device (B). In both measurements, the dc bias current $I$ = 0. The solid lines in both figure 2(a) and (b) are a fit to the data using the functional form predicted by the Ginzburg-Landau (GL) theory for temperatures close to the critical temperature (equation (1)). $L_K(T = 0)$ and $T_c$ were used as fitting parameters. The critical temperature extracted using the GL fit to the data for temperatures near $T_c$ is close to the measured value of $T_c$ for both devices. The value of $L_K(T = 0)$ determined from fitting the measured data near $T_c$ does not agree with the measured value of $L_K$ at lower temperature. This is not unexpected, as equation (1) is a GL prediction that is only valid near $T_c$. The dashed lines show the fit to BCS theory (equation (6)), with $T_c$ as a fitting parameter. The BCS expression fits the data well over the entire temperature range, with fit values of $T_c$ that are in good agreement with the measured values of $T_c$. The values of $L_K(0)$ used in the BCS fits were calculated directly from equation (6) using $\Delta(0) = 1.76k_BT_c$; the calculated values are $L_K(0)$ = 22 nH for the Nb device and $L_K(0)$ = 99 nH for the NbN device.

In figure 2(c), the same data are plotted versus $T/T_c$ by normalizing $L_K$ by $L_K(T_{min})$, where $T_{min}$ is the lowest temperature at which the kinetic inductance was measured. As can be seen, the functional form of the temperature dependence of $L_K$ is very similar for both Nb and NbN nanowires throughout the measured temperature range. Note that we could not measure as close to $T_c$ in the NbN device as in the Nb device because $\Delta T_c$ was larger in the NbN device. This may be because the thicker Nb nanowire has a more uniform cross-section and because of the greater effect of fluctuations at higher temperature.

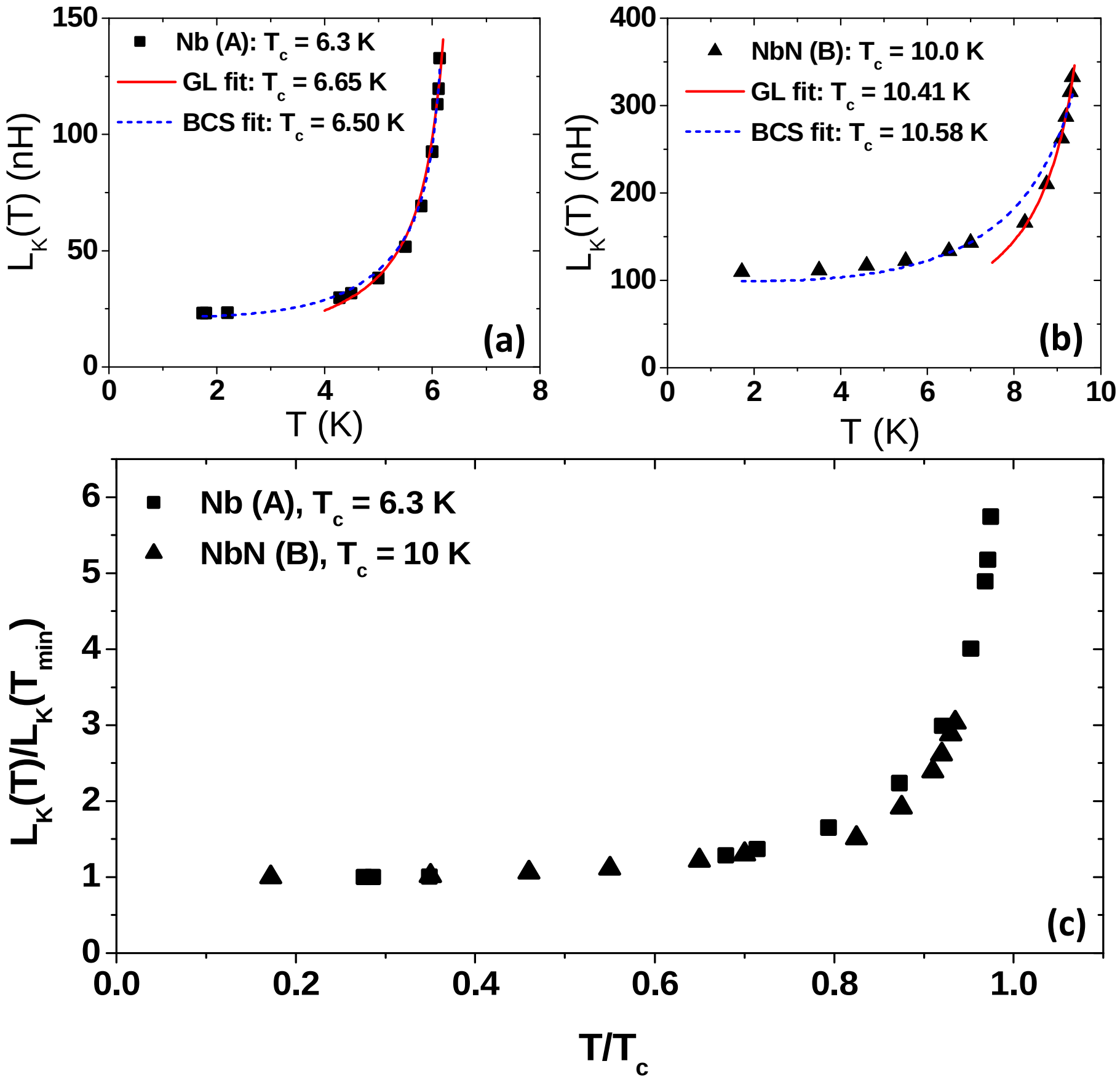

Figure 2. (a) $L_K$ versus $T$ for Nb device (A) with $l$ = 500 μm; (b) $L_K$ versus $T$ for NbN device (B) with $l$ = 105 μm; (c) $L_K/L_K(T_{min})$ versus $T/T_c$ for both devices.

In figure 3, we plot measurements of the dependence of the kinetic inductance on the dc bias current $I$. Figure 3(a) is a plot of $L_K$ versus $I$ for device (A) at $T$ = 2.6 K, and figure 3(b) is a plot of $L_K$ versus $I$ for device (B) at $T$ = 2.5 K. The dashed lines show the fits to BCS theory in the zero temperature limit, found from using equations (11) and (8) in equation (7). The critical current $I_c$ is used as a fitting parameter. The fits result in values of $I_c$ that are approximately 6% larger than the measured $I_c$ for NbN, and approximately 38% larger for Nb. One possible source of this discrepancy is the application of a zero temperature result at finite temperature. We are further from the zero temperature limit in Nb, for which we measure at $T/T_c$ = 0.41, than for NbN, for which we measure at $T/T_c$ = 0.25.

The discrepancy may also be due in part to the measured $I_c$ being reduced by the presence of constrictions, or local regions of reduced critical current, in the nanowire. Using careful measurements of the dc current-voltage characteristics of the nanowires, we can distinguish between devices that are significantly constricted and those that are reasonably unconstricted by the magnitude of the measured critical current and by the presence of extra structure in the current-voltage curves of constricted devices.[17] The current dependence of the kinetic inductance can also be used to screen for constricted devices, as described in reference 15. Hence, while there may be a small reduction in the measured $I_c$ due to local inhomogeneities in the nanowire, we do not believe these devices are significantly constricted. Another possible source of a reduced $I_c$ is the presence of vortex depinning, which can occur at bias currents below the depairing current for wire widths > $4.4\xi$, where $\xi$ is the superconducting coherence length ($\xi \approx$ 10 nm in Nb and $\xi \approx$ 5 nm in NbN).[21]

We also fit the data of figure 3 (a) and (b) to the prediction of GL theory (not shown), which is obtained by numerically solving equations (3) and (4). This yields good agreement with the measured data, similar to the BCS fits, except with fit values of $I_c$ that are approximately twice as large as the measured values. We note, however, that because GL theory is only valid for $T$ near $T_c$, this discrepancy is not surprising.

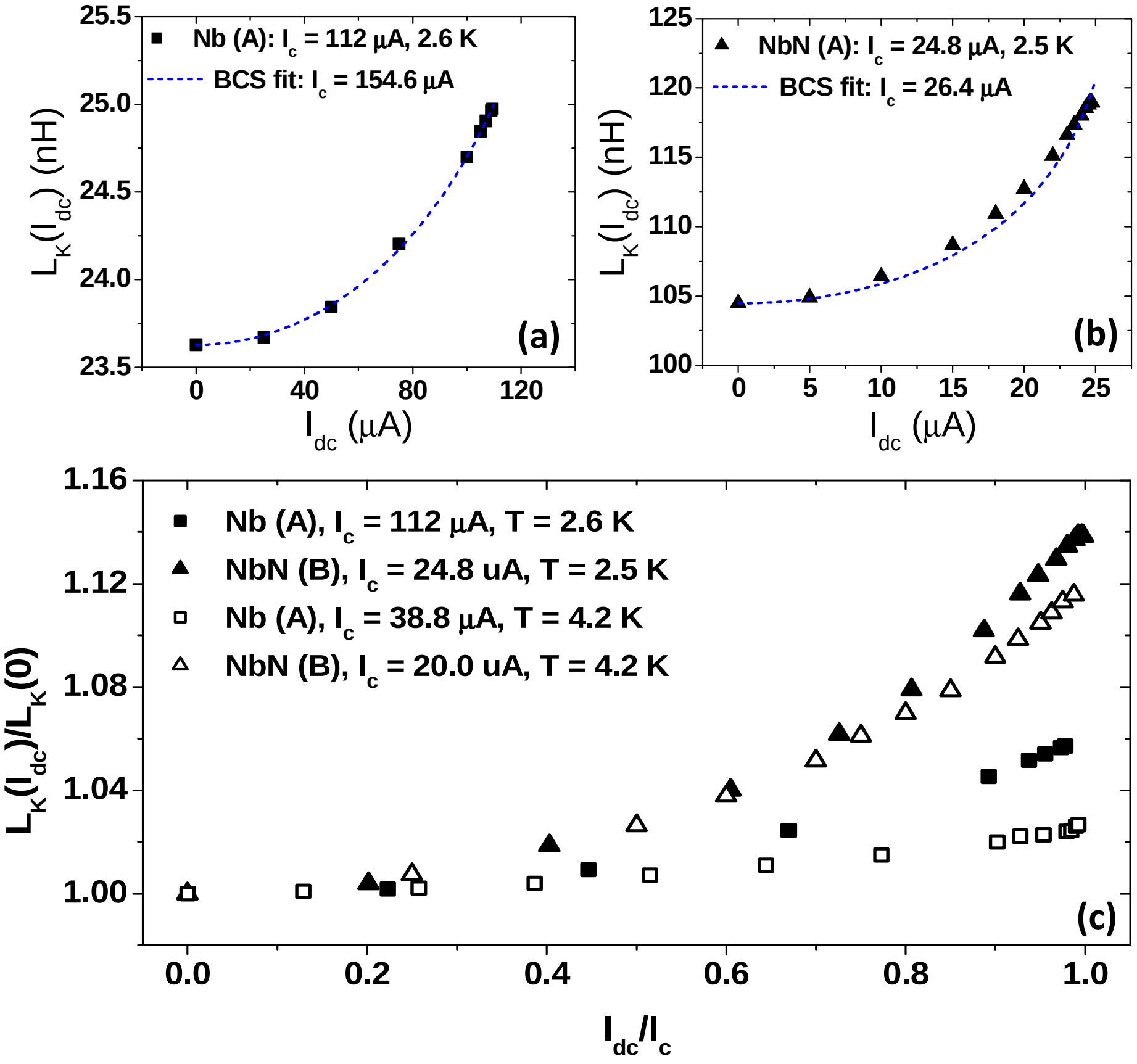

Figure 3. (a) $L_K$ versus $I$ for Nb device (A) at T = 2.6 K; (b) $L_K$ versus $I$ for NbN device (B) at T = 2.5 K; (c) $L_K/L_K(I)$ versus $I/I_c$ for both devices at several temperatures; see text for the fitting procedure.

At higher temperatures ($T > 0.5T_c$) , the inductance versus current in the NbN device displays an unexpected non-monotonicity for bias currents very close to the measured critical current. This is shown in figure 4, where $L_K$ versus $I$ is plotted for temperatures of 4.2 K, 5.0 K, 6.0 K, and 7.0 K. This effect is not seen in the Nb device, which may be due to the measured $I_c$ being further from the true depairing current in this device. Similar behavior has previously been observed in measurements of the kinetic inductance in wider, two-dimensional NbN strips.[12,22,23] In reference 12, this behavior was attributed to thermally-induced vortex motion. In reference 22, this effect was attributed to the thermal fluctuations in intrinsic Josephson junctions formed at grain boundaries in the film. While we presume that this is related to fluctuation effects near the superconducting transition, we note that a quantitative theory for this phenomenon is presently lacking.

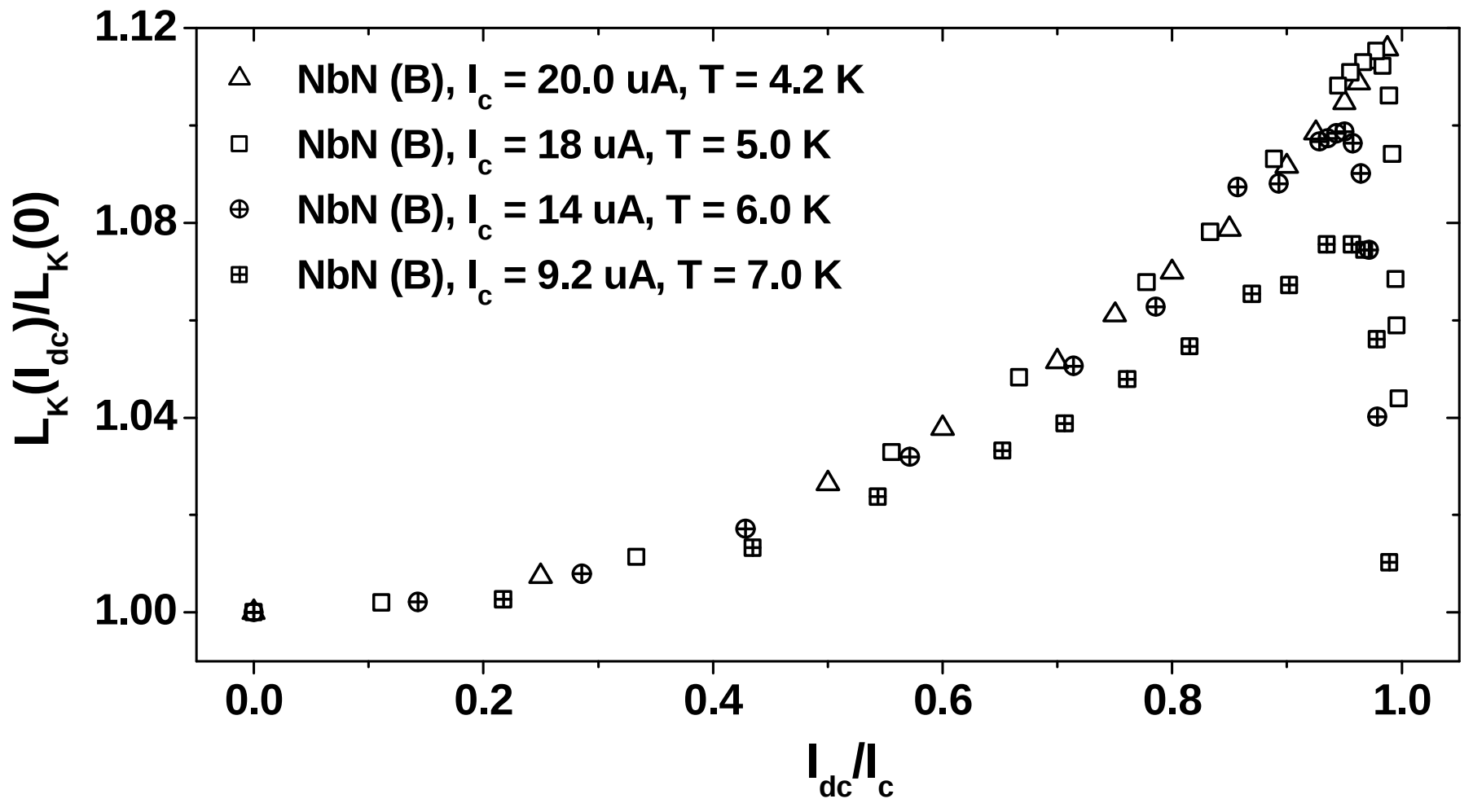

Figure 4: Plot of $L_K$ versus $I$ for NbN device (B) at temperatures of 5.0 K, 6.0 K, and 7 K. The decrease in $L_K$ observed for $I \rightarrow I_c$ at higher temperatures.

The envisioned applications of these devices are in the microwave frequency range. Determination of the self-resonant frequency of these devices is important, as the nanowires are only predominantly inductive for frequencies well below the fundamental self-resonance. It was not possible to measure the self-resonance of the present devices because of parasitic reactance in the electrical coupling structure. We have therefore simulated the self-resonance of these nanowire meander structures using the Sonnet electromagnetic software package.[24] We simulated a 50 Ω microstrip transmission line on a sapphire substrate. The microstrip is interrupted by a nanowire meander with an inductance determined from the measured data at $T$ = 2.5 K or 2.6 K and low bias current. For simplicity, the microstrip width $w$ was chosen to be the same as the overall width of the nanowire meander, and the substrate thickness $d_{sub}$ was chosen to achieve a 50 Ω microstrip impedance ($w \approx d_{sub}$). In figure 5(a), we plot the magnitude of the voltage transmission coefficient $|S_{21}|$ for a microstrip with the 10 x 10 $\mu m^2$ Nb meander ($L_K$ = 22 nH), and in figure 5(b) we plot $|S_{21}|$ for a microstrip with the 5 x 5 $\mu m^2$ NbN meander ($L_K$ = 105 nH). For comparison, we also plot in each figure $|S_{21}|$ calculated for a 50 Ω transmission line interrupted by an ideal lumped element inductor with the same inductance.

The Nb device displays its fundamental self-resonance at approximately 56 GHz, while the NbN device displays its fundamental self-resonance at approximately 37 GHz. Simulations of the same nanowire meanders in a coplanar waveguide geometry yielded very similar results. For each material, the resonant frequency scales inversely with the nanowire length, and hence with the total inductance. The lowest frequency resonance is consistent with a half-wavelength standing wave on the length of the nanowire. This is confirmed by simulating the current density at the first resonance frequency, which is shown for each device in the insets of figure 5. For both devices, a current node (voltage anti-node) is seen at the center of the device, and a current anti-node (voltage node) is seen at each end. The NbN nanowire, although much shorter than the Nb nanowire, has a lower fundamental self-resonant frequency because the inductance per unit length in NbN is much larger, resulting in a slower propagation velocity. Beyond the first resonance, both nanowires exhibit higher order resonances that are not simply integer multiples of the fundamental resonance frequency. This indicates that the simple picture of the nanowire as a section of high impedance transmission line is no longer valid beyond the first resonance, and the behavior at these higher frequencies is more complex.

We also simulated the effect of different nanowire widths. The kinetic inductance should scale as the inverse of the nanowire width. For a fixed length and a fixed spacing between adjacent sections of the meander, the fundamental self-resonant frequency scales approximately as the square root of the wire width, which is consistent with a standing wave resonance. For example, reducing the linewidth of the $l$ = 105 μm NbN nanowire from 130 nm to 50 nm (keeping the same length) increases the inductance to 273 nH and decreases the fundamental self-resonant frequency to 26 GHz. Thus, for a specific material, the highest self-resonant frequency for a given inductance is obtained with the smallest nanowire cross-section.

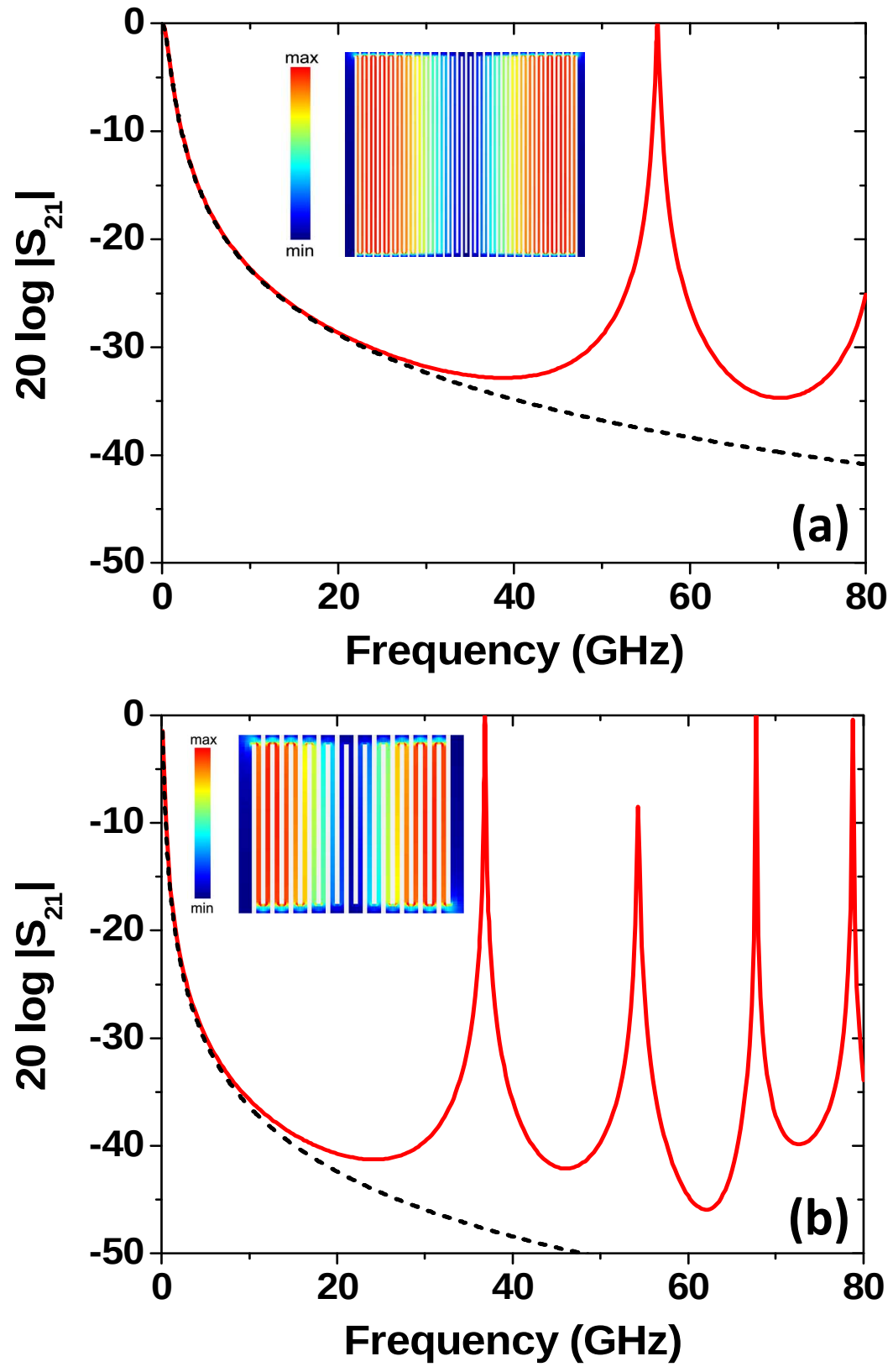

Figure 5. Simulated voltage transmission coefficient $|S_{21}|$ of a 50 Ω microstrip transmission line interrupted by a nanowire meander. (a) 10 x 10 μm$^2$ Nb nanowire meander with $l$ = 500 μm and $L_K$ = 22 nH (solid line), along with the result for an ideal 22 nH inductor (dashed line). (b) 5 x 5 μm$^2$ NbN nanowire meander with $l$ = 105 μm $L_K$ = 105 nH (solid line), along with the result for an ideal 105 nH inductor (dashed line). Insets show the simulated magnitude of the current density for each device at the first resonance.

**4. Conclusions**

We have studied the kinetic inductance of superconducting Nb and NbN nanowires fabricated in a compact meander geometry. These devices combine zero dc resistance with very large inductance, with a number of potential applications in high frequency superconducting circuits. While NbN offers greater inductance per unit length, Nb may be easier to integrate with other superconducting circuit elements. In both materials, $L_K$ can be tuned by more than a factor of 3 by changing the operating temperature, while $L_K$ depends only weakly on the bias current. The measured temperature and current dependence of $L_K$ display reasonable agreement with predictions from BCS theory. Electromagnetic simulations show that the intrinsic self-resonant frequency is approximately 56 GHz for the Nb device ($L_K$ = 22 nH, $l$ = 500 μm), and approximately 37 GHz for the NbN device ($L_K$ = 105 nH, $l$ = 105 μm). The fundamental self-resonance corresponds to a half-wavelength standing wave along the length of the nanowire, where the nanowire acts as a slow-wave, high-impedance transmission line.

**Acknowledgements**
This work was supported by NSF grants ECS-0622035 and DMR-0907082, Yale University, and an NSF Graduate Fellowship (to A.J.A.). L.F. acknowledges partial support from CNR-Istituto di Cibernetica, Pozzuoli, Italy. We thank R. Cristiano and A. Casaburi of CNR-Istituto di Cibernetica for providing the NbN nanowire devices used in this work.